\documentclass[12pt]{article}

\usepackage{latexsym,amssymb,amsmath}

\def\a{\alpha}
\def\b{\beta}

\def\s{\sigma}
\def\u{\tau}

\def\P{\Phi}
\def\q{\quad}

\def\RR{\mathbb{R}}
\def\CC{\mathbb{C}}

\def\sgn{\mathop{\hbox{\rm sgn}}}

\def\Kc{{\cal K}}
\def\Dc{{\cal D}}
\def\Sc{{\cal S}}
\newcommand{\erf}{\mbox{erf}}
\newcommand{\Int}{\int\limits}
\newcommand{\Leq}{\leqslant}
\newcommand{\Geq}{\geqslant}

\newcommand{\Section}[1]{\section{#1}\setcounter{equation}{0}}

\title{
On the Nonlinear Dynamical Equation\\
in the p-adic String Theory
}
\author{
\textsf{V.S.~Vladimirov}\\
\emph{Steklov Mathematical Institute,}\\
\emph{Gubkin St.8, 119991 Moscow, Russia}\\
\emph{email: vladim@mi.ras.ru}\\
\textsf{and}\\
\textsf{Ya.I.~Volovich}\\
\emph{Physics Department, Moscow State University}\\
\emph{Vorobievi Gori, 119899 Moscow, Russia}\\
\emph{email: yaroslav@aylabs.com} }

\date{}

\begin{document}

\maketitle

\begin{abstract}
In this work nonlinear pseudo-differential equations with the infinite
number of derivatives are studied.
These equations form a new class of equations which initially appeared
in p-adic string theory. These equations are of much interest in
mathematical physics and its applications in particular in string
theory and cosmology.

In the present work a systematical mathematical investigation of
the properties of these equations is performed.
The main theorem of uniqueness in some algebra of tempored distributions
is proved. Boundary problems for bounded solutions are studied,
the existence of a space-homogenous solution for odd p is proved.
For even p it is proved that there is no continuous solutions
and it is pointed to the possibility of existence of discontinuous solutions.
Multidimensional equation is also considered and its soliton and q-brane
solutions are discussed.

\end{abstract}

\Section{Introduction}

Recently in works on p-adic and then in real string theories a certain
class of nonlinear equations which involve infinite number of derivatives
is started to be explored \cite{BFOW}-\cite{AJK},
on $p$-adic mathematical physics please see \cite{VVZ,Kh}.
Exploration of this new class of equations is of much interest
in mathematical physics and in the present work a systematical
mathematical investigation of the properties of these equations
is performed.

In string field theory \cite{GSW} the problem of building
dynamics has two important specifics as compared to the
same problem in the local field theory. First of all, string field
theory describes a set of infinite number of local fields.
On the other hand, interaction of the fields in this set is
nonlocal, in the cense that corresponding equations of motion
contain infinite number of derivatives.

The problem of building classical solution which interpolates between
vacua is related with the possible applications in cosmology,
in particular in \cite{Sen} it is proposed to
identify the inflaton field with the tachyon matter in bosonic string
theory.

In $p$-adic string model for the scalar tachyon field it
appeared a new equation of motion -- nonlinear pseudo-differential
equation of the form \cite{BFOW,FO} (please see also \cite{VVZ,BF}
and references there in)
\begin{equation}
\label{padic}
p^{\frac{1}{2}\square}\P=\P^p,
\end{equation}
where
$$
\square= \frac{\partial^2}{\partial_t^2}
        -\frac{\partial^2}{\partial_{x_1}^2}-\cdots
        -\frac{\partial^2}{\partial_{x_{d-1}}^2}
$$
is a D'Alamber operator and $p$ is a prime number, $p=2,3,\ldots$
Although originally in $p$-adic string model $p$ is a prime number
it is still interesting to consider $p$ as an arbitrary integer greater
than one, we will follow this definition of $p$ in the present work.
From the physical point of view only real valued solutions of (\ref{padic})
are interesting, so we  will consider only this type of solutions.
More general equations and systems of equations were obtained and
explored in \cite{AJK,VolYa}.

Let us remind the major steps to obtain a $p$-adic string \cite{VVZ}.
It is well known, that if in the string theory one considers a
tachyon scattering Ve\-ne\-zia\-no's amplitude \cite{GSW} will be obtained,
which could be written in the terms of beta-function on the real numbers filed.
If we replace this beta function with the corresponding $p$-adic
beta function we get the tachyon scattering amplitude in the $p$-adic
string \cite{BFOW,VVZ}.
The regularized adelic formulas for the Veneziano's amplitudes are
obtained in \cite{VS1}.

Equation (\ref{padic}) in the case $p=2$ has two vacuum solutions: $\P_0=0$ and
$\P_0=1$. In the recent work \cite{MolZw} it was performed an investigation
of existence of a solution of (\ref{padic}) which interpolates between these
vacua. It was shown that such monotonic solutions do not exists.
In this work we will prove some more general theorem that states
that there is no even non-monotonic solutions for any even $p$.
In \cite{AJK,VolYa} the same problem for the tachyon in fermionic
string was numerically studied. Here we prove a theorem of
the existence of space-homogenous solution of (\ref{padic}) for
any odd $p$ which interpolates between vacua $\P_0=-1$ and $\P_0=1$.

This work is organized as follows. In the section \ref{setting}
we describe a mathematical problem setup.
In the section \ref{unit-thrm}
the main uniqueness theorem is proved in the ${\tilde \Sc}'_+$
algebra of distributions.
In the section \ref{lim-sol} boundary problems for the limited solutions
of (\ref{padic}) are discussed.
The theorem of existence of space-homogeneous solution
in the case of odd $p$ is proved.
For the case of even $p$ it is proved the lack of a continuous solutions
interpolating between two vacua and it is pointed to the possibility
of the existence of discontinuous solutions.
In the section \ref{multidim} multidimensional and $q$-brane solutions
are discussed.

\Section{Problem Setup}
\label{setting}

In the simplest case of $d=1$ the equation (\ref{padic}) writes as follows
\begin{equation}
\label{1.2}
p^{\frac{1}{2}\partial_t^2}\P=\P^p.
\end{equation}
Let us give the equation (\ref{1.2}) a rigorous meaning.
The equation (\ref{1.2}) is a formal form of a nonlinear pseud-differential
equation with the symbol $e^{-\frac12\xi^2\ln p}$,
\begin{equation}
\label{1.3}
{1\over{2\pi}}\int_{-\infty}^\infty\tilde\P(\xi)\exp(-\frac{1}{2}\xi^2\ln p-it\xi)d\xi=\P^p(t),
\end{equation}
where $\tilde \P (\xi)$ -- it a Fourier transform of a function (distribution) $\P(t)$,
$$
\tilde \P(\xi)=\int_{-\infty}^\infty\P(t)e^{i\xi t}dt.
$$
(Here we use a theory of Fourier transforms of distributions from
the $\Dc'$ class. Fourier transforms of this distributions
are analytical functionals from the space of tempored
distributions $Z'$\cite{GelShil}.)

If we are searching for a solution in the space of tempored
distributions $\Sc'$, then (\ref{1.3}) is equivalent
to the following nonlinear integral equation
\begin{equation}
\label{1.4}
\int_{-\infty}^\infty\P(\u)H[(t-\u)^2]d\u=\P^p(t),
\end{equation}
where the kernel $H[(t-\u)^2]$ is given by
$$
H(t^2)={1\over{2\pi}}\int_{-\infty}^\infty\exp(-\frac{1}{2}\xi^2\ln p-it\xi)d\xi=
$$
\begin{equation}
\label{H-def}
={1\over\sqrt{2\pi\ln p}}\exp \Bigr (-{{t^2}\over{2\ln p}}\Bigl ), \q
  \int_{-\infty}^\infty H(t^2)dt=1,
\end{equation}
which is equivalent (in the terms of Fourier transforms) to the equation with convolutions
\begin{equation}
\label{1.6}
p^{-\frac{1}{2}\xi^2}{\tilde \P}(\xi)={1\over{(2\pi)^{p-1}}}
  (*{\tilde \P})^p (\xi),
\end{equation}
where
$$
(*{\tilde \P})^p(\xi)=\underbrace{(\P*\P*\cdots*\P)}_{\mbox{convolution }(p-1)\mbox{ times}}(\xi).
$$

The left hand side of (\ref{1.4}) is the value of a functional $\P(\u)$
on the test function $H[(t-\u)^2]\in S$, i.e. it is a convolution
$$
(\P*H)(t)=(\P(\u),H[(t-\u)^2]).
$$
The right hand side of (\ref{1.4}) has meaning, if it is considered
in the product algebra of distributions ${\tilde \Sc}'_+\subset \Sc'$.
Now the rigorous meaning of (\ref{1.4}) in the algebra ${\tilde \Sc}'_+$ is
given by
\begin{equation}
\label{1.7}
(\P(\u),H[(t-\u)^2])=\P^p(t).
\end{equation}
Let us remind, that ${\tilde \Sc}'_+$ is a Fourier transform of tempored
distributions with the support on the half-axis $[0,\infty)$.
The algebra ${\tilde \Sc}'_+$ is isomorphic to the convolution algebra
of the boundary values of holomorphic functions $f(z)$, $z=t+iy$
in the upper half-plane $y>0$, which satisfy the following bound condition
\cite{VSComp}
\begin{equation}
\label{1.8}
|f(t+iy)|\Leq C{{1+|z|^\a}\over{y^\b}}, \q y>0
\end{equation}
for some $C>0, \a\Geq 0$ and $\b\Geq 0.$

\Section{The Main Uniqueness Theorem}
\label{unit-thrm}

In this section we will prove the {\it uniqueness} of the solution of
the equation (\ref{1.4}) (more precisely of the equation (\ref{1.7}))
using methods of axiomatic quantum field theory.

{\bf Theorem 1.} {\it Let $\P(t)$ be a real-valued solution of (\ref{1.7})
from the ${\tilde \Sc}'_+$ algebra, then}
\begin{equation}
\label{2.1}
\P(t)=
 \begin{cases}
  0, \hbox{ or } \pm 1, \q &p-\hbox{ odd,} \\
  0, \hbox{ or } 1, \q     &p-\hbox{ even.}
 \end{cases}
\end{equation}

{\bf Proof.}
Let $\P\in{\tilde \Sc}'_+$ be a real-valued solution of (\ref{1.7}).
Then $\P(t)$ is a boundary value in $\Sc'$ of the function $\P(z)$,
$z=t+iy$, which is holomorphic in the upper half-plane $y>0$ and satisfies
the bound condition (\ref{1.8}).

On the other hand from the equation (\ref{1.7}) we have that a distribution
$\P^p(t)$ could be analytically continued to the whole complex plane $z\in\CC$
\begin{equation}
\label{2.2}
\P^p(z)=C_1(\P(\u),e^{-\s(z-\u)^2)}), \q C_1=(2\pi\ln p)^{-1/2}, \q \s=(2\ln p)^{-1}
\end{equation}
and satisfies the bound condition
\begin{equation}
\label{2.3}
|\P^p(z)|\Leq C'(1+|z|)^{2m}e^{\s y^2}, \q z\in\CC,
\end{equation}
where $m$ is the order of distribution $\P$ and constant $C'>0$.

Let us prove the above statement.
The fact that the right hand side of the inequality (\ref{2.2})
is an entire function and its boundary value when $y\to 0$ is equal
to the right hand side of the equation (\ref{1.7}) and thus
equals to the function $\P^p(t)$ could be obtained using
standard methods \cite{VSComp}.

Let us prove the bound condition (\ref{2.3}). From (\ref{2.2}) we get the bound
$$
|\P^p(z)|\Leq C_1\parallel\P\parallel_{-m}\parallel e^{-\s(z-\u)^2}\parallel_m\Leq
$$
$$
C_2\max_{0\Leq j\Leq m}\sup_\u (1+|\u|^j)|{d^j\over{d\u^j}}e^{-\s (z-\u)^2}|\Leq
$$
$$
C_3\sup_\u (1+|\u|^m)(1+|z-\u|^m)|e^{-\s (z^2-2z\u+\u^2)}|\Leq
$$
\begin{equation}
\label{2.4}
C_3(1+|z|^m)e^{\s(y^2-t^2)}\sup_\u (1+|\u|^m)e^{-\s(\u^2-2t\u)}.
\end{equation}
Let us prove the following bound condition
\begin{equation}
\label{2.5}
\sup_\u (1+|\u|^m)e^{-(\u^2-2t\u)}\Leq C_4(1+|t|^m)e^{\s t^2}.
\end{equation}
It is easily seen that the bound (\ref{2.5}) holds in the case $|t|\Leq 1$.
Let us prove this for the case $|t|\Geq 1$. Denoting
$$
f(\rho,|t|)=\rho^m e^{-\s(\rho^2-2\rho|t|)}, \q \rho\Geq 0.
$$
we get
\begin{equation}
\label{2.6}
\sup_\u |\u|^m e^{-\s(\rho^2-2\rho |t|)}\Leq\sup_\u f(\u,|t|)=f(\rho_0,|t|), \q |t|>1.
\end{equation}
Denoting by $\rho_0$ the point where $f(\rho,|t|)$ gets its maximum we have
$$
\rho^2-|t|\rho-{m\over{2\s}}=0,
$$
thus
$$
|t|\Leq\rho_0={|t|\over 2}+\sqrt{{t^2\over 4}+{m\over{2\s}}}\Leq|t|+{m\over {\s|t|}}, \q |t|>1.
$$
Substituting this value of $\rho_0$ to the bound (\ref{2.5}), we get
\begin{equation}
\label{2.7}
\sup_\u |\u|^m e^{-\s (\rho^2-2\rho|t|)}\Leq C_5 (1+|t|^m)e^{\s t^2}, \q |t|>1.
\end{equation}
The bound (\ref{2.7}) holds for $m=0$. Thus the bound (\ref{2.5}) holds
for all $t$. From the bounds (\ref{2.5}) and (\ref{2.4}) it follows the
bound (\ref{2.3}), which leads us to the following bound
\begin{equation}
\label{2.8}
|\P(z)|\Leq C_6(1+|z|)^{{2m}/p}e^{{\s y^2}/p}, \q z\in\CC.
\end{equation}
Now let us prove the bound
\begin{equation}
\label{2.9}
|\P(z)|\Leq C_7(1+|z|)^n, \q z\in\CC
\end{equation}
for some $C_7>0$ and $n>0.$

Let us introduce a function
$$
\P_1(z)=
  \begin{cases}
    \P(z), \q y>0,\\
    \bar\P(\bar z), \q y<0.
  \end{cases}
$$
Since the boundary values $\P_1(x\pm i0)=\P(t)$ of the function $\P_1(z)$ are
all the same (the distribution $\P(t)$ is real!), then following the Bogoliubov's
edge of the wedge theorem \cite{VSComp} the function $\P_1(z)$ is entire
and thus $\P_1(z)=\P(z)$, $z\in\CC$.
Following (\ref{2.7}) and (\ref{1.8}) the function $\P_1(z)$ satisfies the bound
\begin{equation}
\label{2.10}
|\P_1(z)|\Leq C{{1+|z|^\a}\over{|y|^\b}}, \q z\in\CC.
\end{equation}
But it also satisfies the bound (\ref{2.8}).
Thus
$$
|\P(z)|\Leq\min\{C{{1+|z|^\a}\over{|y|^\b}},C_6(1+|z|)^{2m/p}e^{\s y^2/p}\},
$$
from and we get the bound (\ref{2.9}).

Using the Liouville theorem from the bound (\ref{2.10}) it follows that the function
$\P(z)$ is a polynomial of the order not greater than $n$,
\begin{equation}
\label{2.11}
\P(z)=\sum_{k=0}^n a_kz^k,
\end{equation}
and thus
\begin{equation}
\label{2.12}
\tilde\P(\xi)=\sum_{k=0}^n2\pi(-i)^k a_k \delta^{(k)}(\xi).
\end{equation}
Substituting (\ref{2.12}) to (\ref{1.6}) we get
\begin{equation}
\label{2.13}
p^{-\frac{1}{2}\xi^2}\sum_{k=0}^n (-i)^k a_k \delta^{(k)}(\xi)=
  \Bigr(*\sum_{k=0}^n (-i)^k a_k \delta^{(k)}\Bigl )^p(\xi).
\end{equation}
A system of distributions $\{\delta^{(k)}, k=0,1,\ldots\}$ is linear
independent, thus from (\ref{2.13}) we get
$$
(-1)^n a_n\delta^{(n)}(\xi)=(-1)^{kp}a_n^p\delta^{(pn)}(\xi).
$$
But $pn>n$ for $n>0$. Thus $a_n=0$. And so on.
As a result we obtain that in (\ref{2.12}) and (\ref{2.11})
$a_k=0, k=1,2,\ldots,n$. Thus (\ref{2.12}) and (\ref{2.11}) take the form
$\tilde\P(\xi)=2\pi a_0\delta (\xi)$ and $\P(t)=a_0$.
Now using (\ref{1.6}) we get that all possible values of a constant
$a_0$ are $0$ or $1$ for the case of even $p$, and
$0$ and $\pm 1$ for the case of odd $p$. The theorem 1 is proved.

\Section{Boundary Problems for Bounded Solutions}
\label{lim-sol}

The equation (\ref{1.4}) has a rapidly growing solution of the form
$$
\P(t)=\exp{\Bigr ({{\ln p}\over{2(p-1)}}+{{p-1}\over{2p\ln p}}t^2\Bigl )},
$$
this fact could be directly checked using the formula
$$
e^{-\a t^2}*e^{\b t^2}=\sqrt{\pi/(\a-\b)}\exp({{\a\b}/(\a-\b)}t^2), \q \a>\b.
$$
Let us consider bounded solutions $\P(t)$ of the equation (\ref{1.4}).
A question arise: which extra properties will then have the solution?
The function $\P^p(t)$ is a trace for $y=0$ of the entire function
$$
F(z)=\int_{-\infty}^\infty H[(z-\u)^2]\P(\u)d\u, \q z=t+iy\in\CC
$$
(see the proof of the theorem 1 section \ref{unit-thrm}).
This means that $\P(t)$ satisfies the algebraic equation
\begin{equation}
\label{3.1}
\P^p(t)=F(t), \q t\in\RR
\end{equation}
where the function $F(t)$ is bounded and real-valued analytical.
Here one should consider two cases: $p$ is odd and $p$ is even.

In the case of odd $p$ there is a single bounded real-valued solution
of (\ref{3.1}) which is given by $F^{1/p}(t)$.
It is real-valued-analytical where $F(t)\neq 0$.

For even $p$ there discontinuous solutions are possible with discontinuities
of the first type. For example, in the point $t_0$, where $F(t_0)>0$ as a
solution one could take a function of the form
\begin{equation}
\label{3.2}
\P(t)=
  \begin{cases}
    F^{1/p}(t),  &t>t_0,\\
    -F^{1/p}(t), &t<t_0.
  \end{cases}
\end{equation}
On order to omit some exotic solutions of (\ref{3.1}), such as
$$
\P(t)=
  \begin{cases}
    F^{1/p}(t),  &\mbox{if~~} t-\mbox{irrational},\\
    -F^{1/p}(t), &\mbox{if~~} t-\mbox{rational},
  \end{cases}
$$
let us restrict the class of the solutions of the equation (\ref{1.4})
(in the case of even $p$) to the class of bounded real partly-analytical functions.
The resulting discontinuous solutions (\ref{3.2}) of the equation (\ref{3.1})
are real partly-analytical.

Although the problem of finding such points $t_0$ is still left open.
It is possible that these points are real zeros of derivative of
the entire function $F(z)$, i.e. such points $t_0$ that $F'(t_0)=0$.

Now let us consider solutions of (\ref{3.1}) in the vicinities of such points
$t_0$ where $F(t_0)=0$. Each point $t_0$ is a zero of the entire function $F(z)$ and thus
there is such integer $n>0$ and a real-valued-analytical function $F_1(t)\neq 0$
(for even $p$ -- $F_1(t)>0$ and $n$ is even) such that in the vicinity of $t_0$
$$
F(t)=(t-t_0)^n F_1(t).
$$
Thus all possible solutions in the case of even $p$
$$
\P(t)=
  \begin{cases}
    \pm (t-t_0)^{n/p} F_1^{1/p}(t),~~t>t_0,\\
    \pm (t_0-t)^{n/p} F_1^{1/p}(t),~~t<t_0.
  \end{cases}
$$
and a single solution in the case of odd $p$
$$
\P(t)=(t-t_0)^{n/p}F_1^{1/p}(t)
$$
of the equation (\ref{3.1}) are continuous in the point $t_0$
and real partly-analytical in the vicinity of $t_0$.

Summarizing we come to the following conclusion.

{\it In the case of odd $p$ all bounded solutions of the equation (\ref{1.4})
are continuous and real partly-analytical. In the case of even $p$ all
bounded partly-continuous solutions are real partly-analytical,
more over the jumps of the solution $\P(t)$ in the points of discontinuities $t_0$
is equal to either $2\P(t_0)$ or $-2\P(t_0)$.}

Two questions arise:
1). Do there exist real-valued-analytical solutions of the equation (\ref{1.4})?
2). Do there exist discontinuous solutions of the equation (\ref{1.4}) in the case of even $p$?
The answer to these questions could give numerical methods.

{\bf Theorem 2.}
{\it If a solution}\footnote{Not necessarily real solution.}
{\it $\P(t)$ of the equation (\ref{1.4}) is bounded, then it satisfies the bound}
\begin{equation}
\label{3.7}
|\P(t)|\Leq 1, \q t\in\RR.
\end{equation}

{\bf Proof.}
According to our assumption the solution $\P(t)$ is bounded.
Thus there exists a number $M>0$, such that
\begin{equation}
\label{3.8}
\sup_t |\P(t)|=M.
\end{equation}
From the equation (\ref{1.4}) and from (\ref{1.3}) if follows
$$
|\P^p(t)|=|\P(t)|^p=
  |\int_{-\infty}^\infty\P(t)H[(t-\u)^2]d\u|\Leq\int_{-\infty}^\infty |\P(t)|H[(t-\u)^2]d\u\Leq
$$
$$
\sup_t |\P(t)|\int_{-\infty}^\infty H[(t-\u)^2]d\u=M,
$$
thus we obtain the
$$
\sup_t|\P(t)|^p=[\sup_t |\P(t)|]^p=M^p\Leq M,
$$
thus $M\Leq 1$. The theorem 2 is proved.

Other properties of the bounded continuous solutions of the equation (\ref{1.4})
could be found in \cite{MolZw}. In particular it is proved that there is
no monotonically growing solutions $\P(t)$ of the equation (\ref{1.4})
which satisfy the following boundary conditions
\begin{equation}
\label{3.9}
\lim\P(t)=
  \begin{cases}
    0, \q t\to -\infty,\\
    1, \q t\to +\infty.
  \end{cases}
\end{equation}
Let us prove that there is no at all bounded solutions satisfying (\ref{3.9}).
Even more strict statement is true.

{\bf Theorem 3.} {\it There does not exist a nonnegative bounded continuous
solution $\P(t)$ of the following boundary problem for the equation (\ref{1.4})}
\begin{equation}
\label{3.10}
\lim\P(t)=1, \q t\to +\infty
\end{equation}
for some $t_0$
$$
\P(t_0)\Leq 2^{-1/(p-1)}.
$$

{\bf Proof.}
According to the theorem 2, $0<\P(t)<1$.
Using (\ref{3.10}) let us prove that there exists $t_1>t_0$ such that
$\P(t_1)<\P(t_0)$.
Assuming that there is no such $t_1$ we would get the inequality
$\P(t)\Geq\P(t_0)$ for all $t\Geq t_0$.
But then from the equation (\ref{1.4}) it would follow the inequality
$$
\P^p(t_0)=\int_{-\infty}^\infty \P(\u) H[(t_0-\u)^2]d\u >
\P(t_0)\int_{t_0}^\infty H[(t_0-\u)^2]d\u={{\P(t_0)}\over 2},
$$
which contradicts to the inequality (\ref{3.10}).

Let $M$ be a set of $t>t_0$ such that $\P(t)<\P(t_0)$.
As it is seen from what was proved above the set $M$ is not empty.
Let us introduce $T=\sup_{t\in M}t$.
This means that there exists a growing sequence $\{t_k, k=0,1,\ldots\}$
of the points from $M$ such that $t_k\to T$.
If $T<\infty$ then according to the continuousness of the function $\P(t)$,
$\P(t_k)\to\P(T)$ as $k\to\infty$.
But the number $\P(T)\Leq\P(t_0)$ satisfies the inequality (\ref{3.10}).
Thus, as it was proved above, there exists a point $T_1>T$ such that $\P(T_1)<\P(T)\Leq\P(t_0)$,
that contradicts to the definition of the point $T$.
Thus $T=\infty$. But then we get
$$
\P(t_k)<\P(t_0)\Leq 2^{-{1/(p-1)}}<1,
$$
which contradicts to (\ref{3.10}). The theorem 3 is proved.

From the theorem 3, we also get the following consequence:
{\it there does not exist nonnegative bounded continuous solutions of
the boundary problem (\ref{1.4})--(\ref{3.9}).}

{\bf Theorem 4.} {\it There exists a single positive continuous solution
$\P(t)\equiv 1$ of the boundary problem
\begin{equation}
\label{3.11}
\lim\P(t)=
  \begin{cases}
    1, \q t\to -\infty,\\
    1, \q t\to +\infty
  \end{cases}
\end{equation}
for the equation (\ref{1.4}).}

{\bf Proof.}
Indeed $\P(t)\equiv 1$ is a solution of the boundary problem (\ref{1.4})--(\ref{3.11}).
Let us assume that there exists a different solution of the same problem
$0\Leq\P(t)\not\equiv 1$.
As it was proved $0<\P(t)<1$ and thus according to (\ref{3.11}) there exists such $t_0$
that
\begin{equation}
\label{3.12}
0<\P(t_0)=\min_t \P(t) < 1.
\end{equation}
But now from the equation (\ref{1.4}) we get
$$
\P^p(t_0)=\int_{-\infty}^\infty H[(t_0-\u)^2]\P(\u)d\u\Geq\P(t_0),
$$
and thus $\P(t_0)\Geq 1$ which contradicts to (\ref{3.12}).
This contradiction proves the theorem 4.

The problem of existence and uniqueness of the bounded solutions
of (\ref{1.4}), except for the ones described above, is still open.
Although the numerical computations show that for the initial function
\begin{equation}
\label{iter-0}
\P_0(t)=\sgn t
\end{equation}
in the case $p=3$ the iterative process
\begin{equation}
\label{iter}
\P_n(t)=\Bigr (\int_{-\infty}^{\infty}\P_{n-1}(\u)H[(t-\u)^2]d\u\Bigr )^{1/p}, \q n=1,2,\ldots
\end{equation}
rapidly converges \cite{VolYa}.

Here we will prove without using numerical methods
that this iterative process for {\it any} odd $p$
uniformly converges to the solution of a boundary problem
\begin{equation}
\label{3.10-new}
\lim\P(t)=
  \begin{cases}
    -1, &t\to -\infty,\\
    1,  &t\to +\infty
  \end{cases}
\end{equation}
for the equation (\ref{1.4}).

{\bf Lemma}.
If $\P(t)$ is a bounded function on $\RR$ and
\begin{equation}
\label{lemlim}
\lim_{t\to+\infty}\P(t)=1,
\end{equation}
then
\begin{equation}
\label{lemiter}
\lim_{t\to+\infty}\Int_{-\infty}^\infty H[(t-\tau)^2]\P(\tau)d\tau=1
\end{equation}

{\bf Proof}.
From (\ref{lemlim}) and (\ref{H-def}) we obtain (\ref{lemiter}):
\begin{align*}
&\lim_{t\to+\infty}\Int_{-\infty}^\infty H[(t-\tau)^2]\P(\tau)d\tau=\\
&=\lim_{t\to+\infty}\Int_0^\infty H[(t-\tau)^2]\P(\tau)d\tau+
  \lim_{t\to+\infty}\Int_{-\infty}^0 H[(t-\tau)^2]\P(\tau)d\tau=\\
&=\lim_{t\to+\infty}\Int_{-t}^\infty H[u^2]\P(t+u)du+
  \lim_{t\to+\infty}\Int_t^\infty H[u^2]\P(t-u)du=\\
&=\Int_{-\infty}^\infty H[u^2]\lim_{t\to+\infty}\P(t+u)du+
  O(\lim_{t\to+\infty}\Int_t^\infty H[u^2]du)=\Int_{-\infty}^\infty H[u^2]du=1
\end{align*}
In the first integral we used Lebesgue theorem and in the second one we
used the fact that $\P(t)$ is bounded.
The lemma is proved.

{\bf Theorem 5}.
Let $p$ be odd. Then there exists an odd continuous solution of the boundary
problem (\ref{1.4}), (\ref{3.10-new}).

{\bf Proof}.
If a bounded solution exists then following the theorems proved above
it is continuous. Since we are interested in odd solutions then
the problem (\ref{1.4}), (\ref{3.10-new}) is equivalent to the following
boundary problem for the function $\varphi(t)=\P(t\sqrt{2\ln p})$
\begin{equation}
\Int_0^\infty \Kc(t,\tau)\varphi(\tau)d\tau=\varphi^p(t),~~t\Geq 0
\end{equation}
and
\begin{equation}
\label{border}
\lim_{t\to\infty}\varphi(t)=1,
\end{equation}
here
\begin{equation}
\label{K-kernel}
\Kc(t,\tau)=\frac{1}{\sqrt{\pi}}\left[e^{-(t-\tau)^2}-e^{-(t+\tau)^2}\right]
\end{equation}
is a symmetric continuous positive kernel, which becomes equal to zero
when $t=0$ or $\tau=0$. The original function $\P(t)$ is now given by the
following relation
\begin{equation}
\label{P-phi}
\P(t)=
  \begin{cases}
    \varphi(t/\sqrt{2\ln p}),~~t\Geq 0,\\
    -\varphi(t/\sqrt{2\ln p}),~~t<0.
  \end{cases}
\end{equation}

Let us use the iterative process (\ref{iter}), which in the terms of
the corresponding functions $\varphi_n(t)$ will have the form
\begin{equation}
\label{iter-half}
\varphi_n^p(t)=\Int_0^\infty \Kc(t,\tau)\varphi_{n-1}(\tau)d\tau,~~
\varphi_0(t)=1,~~t\Geq0,~~n=0,1,\ldots
\end{equation}

The value of the first iteration $\varphi_1(t)$ is given by
\begin{equation}
\label{iter-1}
\varphi_{1}(t)=\erf(t)^{1/p},
\end{equation}
where the error function $\erf(t)$ is defined by
\begin{equation}
\label{erf-def}
\erf(t) = \frac{2}{\sqrt\pi}\int_0^t \exp(-x^2)dx
\end{equation}
From (\ref{iter-1}), (\ref{erf-def}) it follows that
\begin{equation}
\label{phi1bound}
0\Leq\varphi_1(t)<1=\varphi_0(t),~~\mbox{for all}~t\Geq 0
\end{equation}

Let us prove that the iterative process (\ref{iter-half}) is uniformly
bounded and uniformly convergent.

Please note that using the fact that $\varphi_0(t)=1$ and (\ref{P-phi})
according to the lemma we get that for all iteration numbers $n\Geq 0$
it holds
\begin{equation}
\label{nlim}
\lim_{t\to\infty}\varphi_n(t)=1
\end{equation}

Let us now prove that for the first and second iterations of
the iterative process (\ref{iter-half}) there holds an inequality
\begin{equation}
\label{phi-neq}
\sigma\varphi_1(t)\Leq\varphi_2(t)\Leq\varphi_1(t),~~t\Geq 0,
\end{equation}
for some $\sigma$ which satisfies
\begin{equation}
\label{sigma-bound}
0<\sigma<1.
\end{equation}

First, let us prove that the second part of the inequality (\ref{phi-neq}) holds.
We have
\begin{equation}
\label{phi-neq-2}
\varphi_2(t)^p=\Int_0^\infty \Kc(t,\tau)\varphi_1(\tau)\Leq
 \Int_0^\infty \Kc(t,\tau) d\tau=\erf(t)=\varphi_1(t)^p,
\end{equation}
where we used (\ref{phi1bound}).

Now let us prove that the first part of the inequality (\ref{phi-neq}) holds.
In the case $t=0$ the inequality (\ref{phi-neq}) becomes the equality and thus
holds. We are left now only with first part the inequality (\ref{phi-neq})
in the case of strictly positive $t$.

Let us consider a function $f(t)$ defined as
$$
f(t)=\frac{\varphi_2(t)^p}{\varphi_1(t)^p}
$$
The function $f(t)$ is continuous, positive, and, according to (\ref{phi-neq-2}), $f(t)\Leq 1$.

Let us compute the limit in the point $t=0$. We have
\begin{equation}
\label{dphi1}
\frac{d}{dt}\varphi_1^p(t)=\frac{d}{dt}\erf(t)=\frac{2}{\sqrt\pi}e^{-t^2},~~
  \left. \frac{d}{dt}\varphi_1^p(t)\right|_{t=0}=\frac{2}{\sqrt\pi}
\end{equation}
and
\begin{equation}
\frac{d}{dt}\varphi_2^p(t)=\Int_0^\infty \frac{d}{dt}K(t,\tau)\varphi_1(\tau) d\tau=
\end{equation}
$$
 =\frac{2}{\sqrt\pi}
   \Int_0^\infty [-(t-\tau)e^{(t-\tau)^2}+(t+\tau)e^{(t+\tau)^2}]\erf(\tau)^{1/p} d\tau,
$$
thus
\begin{equation}
\label{dphi2}
\left.\frac{d}{dt}\varphi_2^p(t)\right|_{t=0}=
  \frac{4}{\sqrt\pi}\Int_0^\infty e^{-\tau^2}\tau~\erf(\tau)^{1/p}d\tau
\end{equation}
Then from (\ref{dphi1}) and (\ref{dphi2}) we have
\begin{equation}
\label{lim-0}
\lim_{t\to 0}f(t)=2\Int_0^\infty e^{-\tau^2}\tau~\erf(\tau)^{1/p}d\tau<
 2\Int_0^\infty e^{-\tau^2}\tau d\tau=1
\end{equation}
From the other hand according to (\ref{nlim}) we have
\begin{equation}
\label{lim-infty}
\lim_{t\to \infty}f(t)=\lim_{t\to \infty}\frac{\varphi_2(t)^p}{\varphi_1(t)^p}=1
\end{equation}
The limit (\ref{lim-0}) allows us to consider the function $f(t)$ as a continuous
function on $[0,\infty)$, thus using (\ref{lim-infty}) we obtain that
there exists $\delta>0$ such that $f(t)\Geq\delta>0$.
Now $\sigma$ from (\ref{phi-neq}) is given by $\sigma=\delta^{1/p}$.
This proves that the inequality (\ref{phi-neq}) holds.

Using the fact that the kernel $K(t,\tau)$ is positive we can integrate the
inequality (\ref{phi-neq})
\begin{equation}
\label{equal-int}
\sigma\Int_0^\infty \Kc(t,\tau)\varphi_1(\tau)d\tau\Leq
 \Int_0^\infty \Kc(t,\tau)\varphi_2(\tau)d\tau\Leq
 \Int_0^\infty \Kc(t,\tau)\varphi_1(\tau)d\tau
\end{equation}
The inequality (\ref{equal-int}) gives us
\begin{equation}
\label{equal-p}
\sigma\varphi_2(t)^p\Leq\varphi_3(t)^p\Leq\varphi_2(t)^p,~~\mbox{i.e.}~~
\sigma^{1/p}\varphi_2(t)\Leq\varphi_3(t)\Leq\varphi_2(t),
\end{equation}
and so on, we obtain
\begin{eqnarray}
\label{equal-n1-ss-2}
\sigma^{1/p^{n-1}}\varphi_n(t)\Leq\varphi_{n+1}(t)\Leq\varphi_n(t),
\end{eqnarray}
thus
\begin{eqnarray}
\label{equal-n1-ss}
0\Leq\varphi_n(t)-\varphi_{n+1}(t)\Leq\varphi_n(t)(1-\sigma^{1/p^{n-1}})
\end{eqnarray}
From (\ref{phi1bound}) and (\ref{equal-n1-ss-2}) it follows that $0\Leq\varphi_n(t)<1$,
thus
\begin{eqnarray}
\label{req}
|\varphi_n(t)-\varphi_{n+1}(t)|<\frac{-\ln\sigma}{p^{n-1}},~~n\Geq 1
\end{eqnarray}
Here we used the following inequality
\begin{equation}
\label{sigma-neq}
1-\sigma^\alpha < -\alpha\ln\sigma,
\end{equation}
which holds for any $\alpha>0$ and $\sigma\in(0,1)$.
To prove (\ref{sigma-neq}) let us consider a function
$$
f(\sigma)=-\alpha\ln\sigma+\sigma^\alpha-1,
$$
we see that $f(0)\to+\infty$, $f(1)=0$, and the derivative is negative
$$
f'(\sigma)=-\frac{\alpha}{\sigma}+\alpha\;\sigma^{\alpha-1}=
  \frac{\alpha}{\sigma}(\sigma^\alpha-1)<0
$$

Let us note, that according to (\ref{sigma-bound}) the right hand side of the inequality
(\ref{req}) is always positive. From (\ref{req}) it follows the uniform convergence
of the sequence $\varphi_n(t)$ for $t\Leq 0$.

Let us prove now, that the function $\varphi(t)$ defined as
\begin{equation}
\label{solution}
\varphi(t)=\lim_{n\to\infty}\varphi_n(t)=
     \varphi_0(t)+\sum_{n=0}^{\infty}[\varphi_{n+1}(t)-\varphi_n(t)]
\end{equation}
satisfies the boundary condition (\ref{border}).

As is was proved above the series (\ref{solution}) uniformly converges for $0\Leq t<\infty$,
thus, using (\ref{nlim}) and taking the limit $t\to\infty$ we get
$$
\lim_{t\to\infty}\varphi(t)=\lim_{t\to\infty}\varphi_0(t)+
 \lim_{t\to\infty}\sum_{n=0}^{\infty}[\varphi_{n+1}(t)-\varphi_n(t)]=1+\sum_{n=0}^{\infty}[1-1]=1,
$$
that proves the property (\ref{border}), and thus the whole theorem.


It is easily seen that the mirrored function $\P(-t)$ is also the solution of the
equation (\ref{1.4}) with the mirrored boundary conditions (\ref{3.10-new})
$$
\lim_{t\to-\infty}\P(-t)=1
$$

The uniqueness of the described type of solutions is still left open.

\Section{Multidimensional Equations of Motion for $p$-adic String}
\label{multidim}

Following the same concept as in the section \ref{setting}, we see that
the equation (\ref{padic}) is a $d$-dimensional pseudo-differential equation
of the form
$$
{1\over{(2\pi)^{d-1}}}\int_{\RR^d}{\tilde\P}_x(\u,\xi)H[(t-\u)^2]
 \exp(\frac{1}{2}|\xi|^2\ln p-i(x,\xi))d\u d\xi=
$$
\begin{equation}
\label{4.1}=\P^p(t,x),
\end{equation}
where ${\tilde \P}_x(\u,\xi)$ is a Fourier transform with respect to variables
$x=(x_1,\ldots,x_{d-1}),$
$$
|\xi|^2=\xi_1^2+\ldots+\xi_{d-1}^2, \q (x,\xi)=x_1\xi_1+\ldots+x_{d-1}\xi_{d-1}.
$$
The corresponding one-dimensional equation with the variable $x_j=x$
has the form (see also (\ref{1.2}))
\begin{equation}
\label{4.2}
\Psi^p (x)={1\over{2\pi}}\int_{-\infty}^\infty
     {\tilde\Psi}(\xi)\exp(\frac{1}{2}\xi^2\ln p-ix\xi)d\xi.
\end{equation}
It has a soliton solution (see also (\ref{3.1}))
\begin{equation}
\label{4.3}
\Psi (x)=\exp\Bigr ({{\ln p}\over{2(p-1)}}-{{p-1}\over{2p\ln p}}x^2\Bigl)
\end{equation}
and trivial solutions which where pointed out in the theorem 1,
section \ref{unit-thrm}.

Using soliton solutions of (\ref{4.3}) one could construct $q$-brane ($q=0,1,\ldots,d-2$)
soliton solutions of the equation (\ref{4.1}) which do not
depend on $t,x_1,\ldots,x_q$
\begin{equation}
\label{4.4}
\P (x_{q+1},\ldots,x_{d-1})=\Psi (x_{q+1})\Psi (x_{q+2})\ldots\Psi (x_{d-1})
\end{equation}
(please see \cite{GhSen} and references there in).

The physical meaning of the soliton solutions of (\ref{4.4}) is discussed
in many recent works in particular in \cite{BF}-\cite{MolZw},\cite{GhSen}-\cite{VolYa}.

\centerline{$*$}
\centerline{$* \q *$}

\Section{Acknowledgments}

Authors are grateful to I.Ya.~Aref'eva and I.V.~Volovich for the fruitful discussions.

This work was done with partial financial support by the Russian president's grant
for the leading scientific schools NSh-1542.2003.1.
Ya.V. is partly supported by the grants RFFI-02-01-01084 and RFFI-MAS-03-01-06466.

\end{document}